\documentclass[times, twocolumn, trackchanges]{aastex63}
\usepackage{graphicx, graphics}
\usepackage{CJK}
\usepackage{bbm}
\definecolor{cyan(process)}{rgb}{0.0, 0.72, 0.92}
\usepackage{amsmath, amssymb, bm}
\usepackage[nodayofweek]{datetime}
\newdateformat{monthyearday}{%
  \THEYEAR\ \monthname[\THEMONTH] \twodigit{\THEDAY}}

\newcommand{\Qphi}{$\mathcal{Q}_\phi$}

\newcommand{\uatnum}[1]{\href{http://vocabs.ands.org.au/repository/api/lda/aas/the-unified-astronomy-thesaurus/current/resource.html?uri=http://astrothesaurus.org/uat/#1}{#1}}
\defcitealias{ren18b}{R18}

\received{2020 May 8}
\revised{2020 July 8} %\monthyearday\today
\accepted{2020 July 8}
\published{2020 July 29}
\acceptjournal{The Astrophysical Journal Letters}

\shorttitle{MWC~758 Spiral Arm Shadow \& Motion}
\shortauthors{Ren et al.}

\begin{document}
\pagenumbering{arabic}
\begin{CJK*}{UTF8}{gbsn}
\title{Dynamical Evidence of a Spiral Arm--Driving Planet\\ in the MWC~758 Protoplanetary Disk}
\author[0000-0003-1698-9696]{Bin Ren (任彬)}\email{ren@caltech.edu}
\affiliation{Department of Astronomy, California Institute of Technology, 1216 East California Boulevard, Pasadena, CA 91125, USA}

\author[0000-0001-9290-7846]{Ruobing Dong (董若冰)}
\affiliation{Department of Physics \& Astronomy, University of Victoria, Victoria, BC, V8P 1A1, Canada}

\author[0000-0003-1520-8405]{Rob G. van Holstein}
\affiliation{Leiden Observatory, Universiteit Leiden, PO Box 9513, 2300 RA Leiden, The Netherlands}
\affiliation{European Southern Observatory, Alonso de C\'ordova 3107, Casilla 19001, Vitacura, Santiago, Chile}

\author[0000-0003-2233-4821]{Jean-Baptiste Ruffio}
\author{Benjamin A. Calvin}
\affiliation{Department of Astronomy, California Institute of Technology, 1216 East California Boulevard, Pasadena, CA 91125, USA}

\author{Julien H. Girard}
\affiliation{Space Telescope Science Institute, 3700 San Martin Drive, Baltimore, MD 21218, USA}

\author[0000-0002-7695-7605]{Myriam Benisty}
\affiliation{Universit\'e Grenoble Alpes, CNRS, IPAG, 38000 Grenoble, France}
\affiliation{Unidad Mixta Internacional Franco-Chilena de Astronom\'{i}a (CNRS, UMI 3386), Departamento de Astronom\'{i}a, Universidad de Chile, Camino El Observatorio 1515, Las Condes, Santiago, Chile}

\author{Anthony Boccaletti}
\affiliation{LESIA, Observatoire de Paris, PSL Research University, CNRS, Sorbonne Universit\'es, UPMC, Univ. Paris 06, Univ. Paris Diderot, Sorbonne Paris Cit\'e, 5 place Jules Janssen, 92195 Meudon, France}

\author[0000-0002-0792-3719]{Thomas M. Esposito}
\affiliation{Department of Astronomy, University of California, Berkeley, CA 94720, USA}

\author[0000-0002-9173-0740]{\'Elodie Choquet}
\affiliation{Aix Marseille Univ, CNRS, CNES, LAM, Marseille, France}

\author[0000-0002-8895-4735]{Dimitri Mawet}
\affiliation{Department of Astronomy, California Institute of Technology, 1216 East California Boulevard, Pasadena, CA 91125, USA}

\author{Laurent Pueyo}
\affiliation{Space Telescope Science Institute, 3700 San Martin Drive, Baltimore, MD 21218, USA}

\author[0000-0002-5823-3072]{Tomas Stolker}
\affiliation{Institute for Particle Physics and Astrophysics, ETH Zurich, Wolfgang-Pauli-Strasse 27, 8093 Z\"{u}rich, Switzerland}

\author{Eugene Chiang}
\affiliation{Department of Astronomy, University of California, Berkeley, CA 94720, USA}

\author{Jozua de Boer}
\affiliation{Leiden Observatory, Universiteit Leiden, PO Box 9513, 2300 RA Leiden, The Netherlands}

\author[0000-0002-1783-8817]{John H. Debes}
\affiliation{Space Telescope Science Institute, 3700 San Martin Drive, Baltimore, MD 21218, USA}

\author{Antonio Garufi}
\affiliation{INAF, Osservatorio Astrofisico di Arcetri, Largo Enrico Fermi 5, 50125 Firenze, Italy}

\author{Carol A. Grady}
\affiliation{Exoplanets and Stellar Astrophysics Laboratory, Code 667, Goddard Space Flight Center, Greenbelt, MD 20771, USA}

\author[0000-0003-4653-6161]{Dean C. Hines}
\affiliation{Space Telescope Science Institute, 3700 San Martin Drive, Baltimore, MD 21218, USA}

\author{Anne-Lise Maire}
\affiliation{Space Sciences, Technologies, and Astrophysics Research (STAR) Institute, Universit\'e de Li\`ege, Li\`ege, Belgium}

\author[0000-0002-1637-7393]{Fran\c{c}ois M\'enard} 
\affiliation{Universit\'e Grenoble Alpes, CNRS, IPAG, 38000 Grenoble, France}

\author[0000-0001-6205-9233]{Maxwell A. Millar-Blanchaer}
\affiliation{Department of Astronomy, California Institute of Technology, 1216 East California Boulevard, Pasadena, CA 91125, USA}

\author[0000-0002-3191-8151]{Marshall D. Perrin}
\author[0000-0003-4845-7483]{Charles A. Poteet}
\affiliation{Space Telescope Science Institute, 3700 San Martin Drive, Baltimore, MD 21218, USA}

\author[0000-0002-4511-5966]{Glenn Schneider}
\affiliation{Steward Observatory, The University of Arizona, Tucson, AZ 85721, USA}

\begin{abstract}%\vspace{-0.4cm}
More than a dozen young stars host spiral arms in their surrounding protoplanetary disks. The excitation mechanisms of such arms are under debate. The two leading hypotheses -- companion-disk interaction and gravitational instability (GI) -- predict distinct motion for spirals. By imaging the MWC~758 spiral arm system at two epochs spanning ${\sim}5$~yr using the SPHERE instrument on the Very Large Telescope (VLT), we test the two hypotheses for the first time. We find that the pattern speeds of the spirals are not consistent with the GI origin. Our measurements further evince the existence of a faint ``missing planet'' driving the disk arms. The average spiral pattern speed is $0\fdg22\pm0\fdg03$~yr$^{-1}$, pointing to a driver at $172_{-14}^{+18}$~au around a $1.9$ $M_\sun$ central star if it is on a circular orbit. In addition, we witness time varying shadowing effects on a global scale that are likely originated from an inner disk. 
\end{abstract}


\section{Introduction}

Spiral arms, spanning from tens to hundreds of au, are found in more than a dozen protoplanetary disks in visible to near-infrared light with high-contrast imaging \citep[e.g.,][]{grady99, muto12, grady13, wagner15, monnier19, garufi20, muroarena20, menard20}. Their origin has profound implications for both planet formation and disk evolution \citep{dong18b, brittain20}. In the companion-disk interaction scenario \citep{kley12, dong15, zhu15, bae16}, the masses and locations of the drivers can be inferred \citep{fung15, dong17}, while in the GI scenario \citep{lodatorice15, dong15b, kratter16}, the disk masses can be constrained. To test the two hypotheses, great effort has been expended to search for faint companions in disks, and to accurately measure disk masses. However, both approaches are notoriously difficult. As a result, only the spiral arms in two systems have been confirmed to be induced by stellar companions (HD~100453: \citealp{rosotti20}; UX~Tau: \citealp{menard20}).

The pattern speed of the spirals provides an exciting route to test  hypotheses: the arms corotate with the driver in the companion scenario, and undergo local Keplerian motion on timescales much shorter than the dynamical timescale in the GI scenario. Long temporal baseline and high resolution imaging are needed to accurately assess the motion of the spirals (\citealp{ren18b}, hereafter \citetalias{ren18b}). By far, there has been no observational attempt to dynamically test the two arm formation and thus motion hypotheses. In this Letter, we image the MWC~758 protoplanetary disk with a $4.71$~yr baseline to investigate the change in the brightness and the motion of the spiral arms between two observations, and dynamically test the two arm motion hypotheses.

Located at $160.2 \pm 1.7$~pc \citep{gaiadr2}, the A8V Herbig star MWC~758 \citep{vieira03} is $10.9_{-1.0}^{+12.0}$~Myr old with an estimated mass of $1.9\pm0.2$ $M_\sun$ \citep{garufi18}\footnote{The previous age estimate of $3.5\pm2.0$ Myr \citep{vieira03} has been updated after \textit{Gaia} DR2 by, e.g., \citet{garufi18} and \citet{vioque18} who also calculated the star mass. Here we adopt the \citet{garufi18} values since their star mass is consistent with the CO line emission measurement by \citet{isella10}.}. It hosts a protoplanetary disk with two prominent spiral arms in near-infrared light (\citealp{grady13, benisty15, reggiani18}; \citetalias{ren18b}; \citealp{wagner19}). The spiral arms have been hypothesized to be driven by GI \citep{dong15b} or a planetary companion \citep{dong15, baruteau19}. High-contrast imaging searches have identified two candidates in the disk \citep{reggiani18, wagner19}, and their nature is still under investigation. Under the planet driver assumption, the inner candidate in \citet{reggiani18} has been ruled out as the arm driver by the motion measurements in \citetalias{ren18b}, unless it is on a highly eccentric orbit \citep{calcino20}.

\section{Observation and Data Reduction}
\begin{figure*}[htb!]
\center
\includegraphics[width=\textwidth]{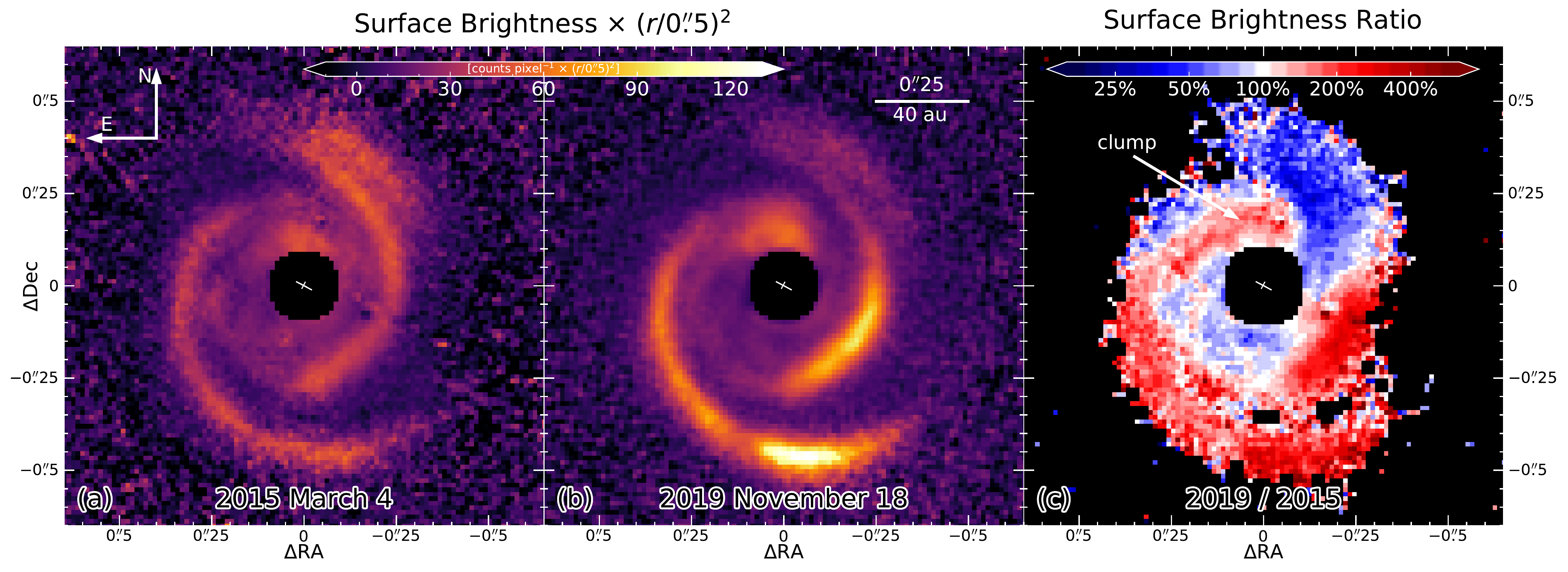}
\caption{VLT/SPHERE observations of MWC~758 spiral arms. (a) and (b) are the 2015 and 2019 scaled surface brightness Stokes \Qphi\ images. (c) is the surface brightness ratio from dividing (b) by (a). In 2019, most of the southern region is more than twice the corresponding brightness of 2015; the northern region is less than half except for a brightened clump. The $184$~mas diameter coronagraph used in 2019 blocks light in the central circular regions. 
}
\label{fig1}
\end{figure*}
We observe MWC~758 at two epochs using the infrared dual-band imager and spectrograph on SPHERE \citep{beuzit19} with the differential polarimetric imaging mode at $Y$-band (1.04~$\mu$m).
The first epoch is on 2015 March 4 under ESO program 60.A-9389(A) (PI: M.~Benisty; \citealp{benisty15}). The second epoch is on 2019 November 18 under ESO program 104.C-0472(A) (PI: B.~Ren). 

In both epochs, the detector integration time is 32~s per frame.  We obtain the 2015 data in field-tracking mode using the apodized Lyot coronagraph
with apodizer APO2, which is optimized for  $5.2\lambda/D$ focal masks, and the 145 mas diameter Lyot mask ALC1 (coronagraph combination name: {\tt N\_ALC\_Y}, inner working angle: IWA $=72.5$~mas, $1$ pixel is $12.25$~mas: \citealp{maire16}). We have $4$ polarimetric cycles. In each cycle, the half-wave plate (HWP) cycles through switch angles $0^\circ$, $22\fdg5$, $45^\circ$ and $67\fdg5$ to measure Stokes $Q$ and $U$. At each HWP position, there are $6$ integrations. The atmospheric seeing (as measured by the differential image motion monitor; DIMM) is $1\farcs08\pm0\farcs17$\footnote{The uncertainties in this Letter are $1\sigma$ unless otherwise specified.} and the coherence time is $3$~ms.  We obtain the 2019 data in pupil-tracking mode to clean and stabilize the diffraction pattern using the apodized Lyot coronagraph using apodizer APO1, which is optimized for $4\lambda/D$ focal masks, and the 185 mas diameter Lyot mask ALC2 (coronagraph combination name: {\tt N\_ALC\_YJH\_S}, IWA $=92.5$~mas). We have 10 polarimetric cycles with HWP switch angles $0^\circ$, $45^\circ$, $22\fdg5$ and $67\fdg5$. There are $2$ integrations at each HWP position. The seeing and coherence time are $0\farcs63\pm0\farcs06$ and $5$~ms, respectively. In our $1$ hour observation blocks, we total $3072$ s on-source integration time in 2015 and $2569$ s in 2019, with $128$ s and $32$ s on-sky time at the end of each observation, respectively.

We reduce the two data sets using the {\tt IRDAP} %\footnote{\url{https://irdap.readthedocs.io}} 
data-reduction pipeline \citep{vanholstein17, vanholstein20} that employs a fully validated Mueller matrix model to minimize reduction bias. We use the images with star polarization subtracted for our analysis. Specifically, we use the \Qphi\ images that show the light polarized parallel or perpendicular to the radial direction from the star, and trace the dust particles on the surface of a disk \citep{monnier19}. We measure that the flux in the central star's point spread function halo in the total intensity images of the 2015 data is $90\%\pm2\%$ of that in the 2019 data. We therefore divide the 2015 \Qphi\ image by $0.9$ to minimize effects from the central star illumination and/or observation conditions.

We scale the surface brightness distribution for the two \Qphi\ images for comparison and analysis. First, we deproject the images to face-on view assuming an inclination of $21^\circ$ and a position angle of $62^\circ$ for the disk \citep{isella10, boehler18}. Next, we compute the mid-plane stellocentric distances ($r$) 
for all pixels. We multiply the value at each pixel by $(r/r_0)^2$, where $r_0=0\farcs5$, to enhance the visibility of features at large distances.
We present the resulting surface brightness maps at the two epochs in Figure~\ref{fig1}, as well as the ratio between the two.
We mark the star location with a white cross, whose longer axis is aligned to the disk major axis.

\begin{figure*}[htb!]
\center
\includegraphics[width=\textwidth]{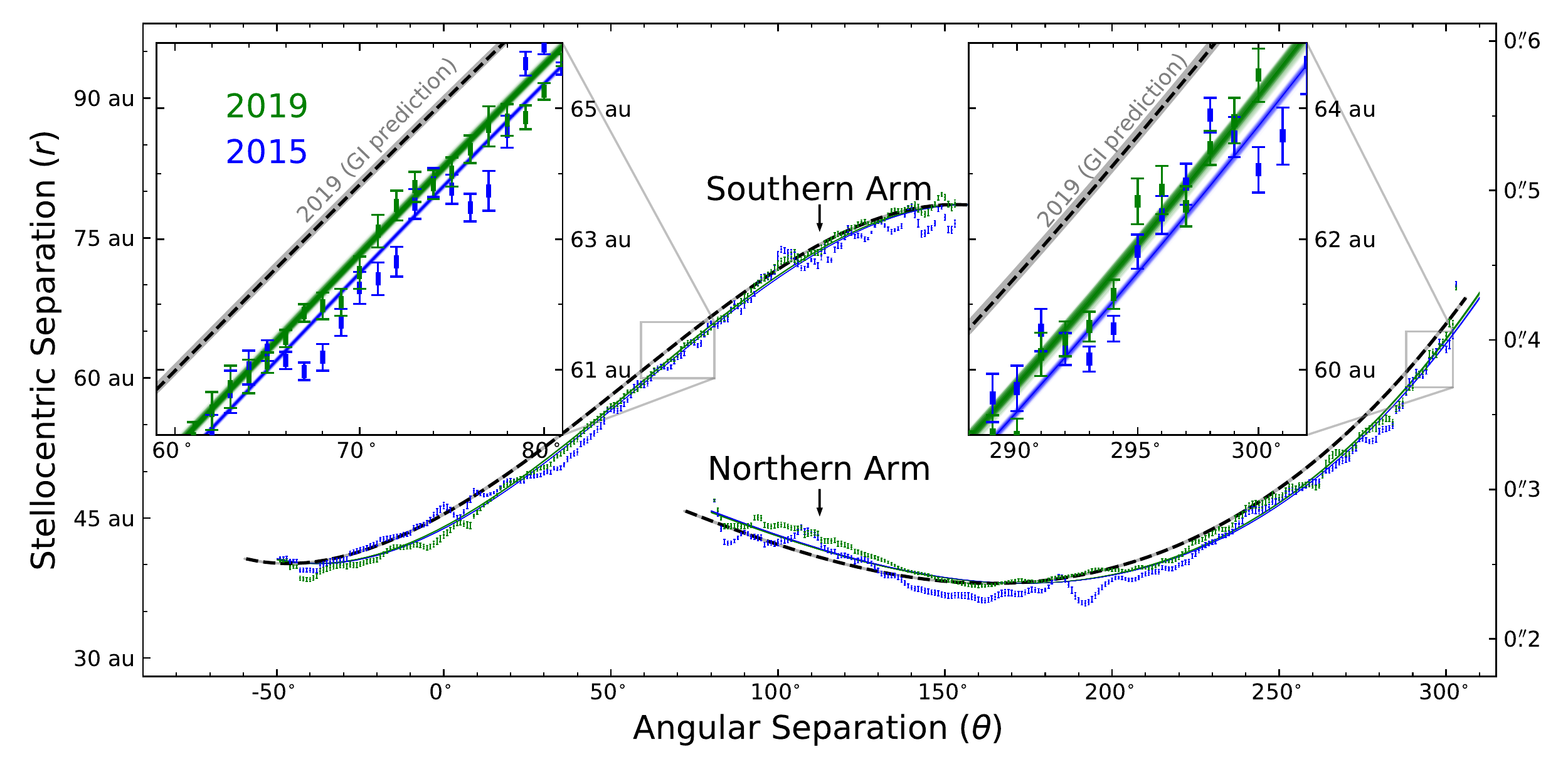}
\caption{Arm location pairs in polar coordinates. The error bars are the peak positions of spiral arms in the form of $(\theta, r)$ pairs in the deprojected version of Figure~\ref{fig1}, where $\theta$ is the counter-clockwise deviation from the northeast semi-major axis. The colored lines are the best fit companion driven model to the observations. The angular offset between the colored lines is $1\fdg04\pm0\fdg14$. The observation does not follow the GI prediction for a $1.9\pm0.2$ $M_\sun$ central star.}
\label{fig-measure}
\end{figure*}

\section{Analysis}
\subsection{Shadowing Effects}
We identify time-varying illumination patterns on a global scale in Figure~\ref{fig1}. Since scattered light probes the disk surface, we interpret the variations as moving shadows. Similar large-scale
shadowing effects have only been observed for TW~Hya in \citet{debes17}.
For MWC~758, the southern region is twice as bright in 2019 than in 2015, while the northern arm on the opposite side (${\sim}1$ o'clock) has dimmed by a factor of two.

In scattered light, the brightness variation at a location may be caused by a change in the shadow casting inner disk structure at the same azimuth, resulting from mechanisms such as the precessing of an inner disk behind the coronagraphic mask \citep{nealon19}
or fluctuations in the dust structure in the inner disk 
arising 
from dust dynamics \citep{stolker17}. We detect a central source degree of linear polarization of $0.50\%\pm0.06\%$ and an angle of linear polarization of $126^\circ\pm3^\circ$ using {\tt IRDAP}, which could originate from an inclined inner disk \citep{vanholstein20}. A clump right outside the north edge of the coronagraph (marked in panel (c) in Figure~\ref{fig1}) has brightened by a factor of two from 2015 to 2019. Given that its brightening coincides with the fainting of the spiral arm tip in the northern region at a larger radii, we hypothesize that the clump may be shadowing the outer disk.

\subsection{Arm Motion}\label{sec-arm-motion-physical}
We deproject the scaled \Qphi\ images to face-on views to measure the location of the spiral arms. For each angle $\theta$, which is defined as the counter-clockwise deviation from the northeast semi-major axis of the disk, we fit a Gaussian profile to its corresponding radial profile to obtain the peak location $r$ with error $\delta r$ using {\tt scipy.optimize.curve\_fit} \citep{scipy19}. We obtain the $(\theta, r)$ pairs with $1^\circ$ step, and present the measurements in Figure~\ref{fig-measure}.

We constrain the morphology and quantify the angular offset between the two epochs for each arm under the two hypotheses. Under different motion mechanisms, a $(\theta, r)$ pair in the first epoch will advance to $(\theta+\Delta\theta, r)$ in the second epoch, where  $\Delta\theta$ is the angular offset between the epochs. On one hand, in the GI scenario, each part of the arm moves roughly on a circular orbit at the local Keplerian velocity on short timescale, $\Delta\theta\propto r^{-3/2}$, and the arms wind up with time
%The arms become more tightly wound with time, and 
as the local pattern speed decreases with increasing stellocentric distance (See, e.g., \citealp{pfalzner03} for the winding up of spiral arms).
On the other hand, in the companion scenario, an entire arm corotates around the star as a rigid body with its driver, and the angular offset between epochs is radius independent. %\added{Under this mechanism, $\Delta\theta$ is independent of $r$.}
%RD: this is really redundant with the previous sentence, i think.
We fit $p$-degree polynomials to the $(\theta, r)$ pairs in both epochs with predicted locations to simultaneously constrain arm morphology and obtain the motion between different epochs, see \ref{math-measure} for the mathematical formalism.

In the GI-induced scenario, if the two arms are undergoing rotation at the local Keplerian speed,  the fitted pattern speed is $(0\fdg058{\pm}0\fdg009){\times}{(0.\!\!''5/r)^{3/2}}$~yr$^{-1}$. To take into account of the $0\fdg08$ true north uncertainty of SPHERE \citep{maire16}, which affects the position angle measurement towards the same direction within each epoch, we first propagate the $0\fdg009$~yr$^{-1}$ measurement uncertainty using the temporal separation between the two epochs, then we combine it with the instrumental true north uncertainty for two observations assuming no correlated noise. Finally, we obtain an uncertainty of $\sqrt{(0.009\times4.71)^2 + 2 \times 0.08^2}/4.71 = 0\fdg03$~yr$^{-1}$, thus the updated motion rate is $(0\fdg06{\pm}0\fdg03){\times}{(0.\!\!''5/r)^{3/2}}$~yr$^{-1}$. This rate corresponds to a central star mass of $0.014_{-0.010}^{+0.018}$ $M_\sun$, two orders of magnitude smaller than the current estimate of $1.56_{-0.08}^{+0.11}~M_\sun$ or $1.9\pm0.2$ $M_\sun$ \citep{vioque18, garufi18}. We thus rule out the GI origin of the spirals at ${>}5\sigma$ levels. Furthermore, the symmetric two-arm morphology in a GI disk in scattered light suggests a disk-to-star mass ratio of $\gtrsim 0.25$ \citep{dong15b}, which corresponds to a high accretion rate. Therefore, the disk would have been dissipated given the age of MWC~758. For illustration, we use the constrained morphology of 2015 spiral arms to predict their locations in 2019, see Figure~\ref{fig-measure}. 

\begin{figure}[hbt!]
\center
\includegraphics[width=0.5\textwidth]{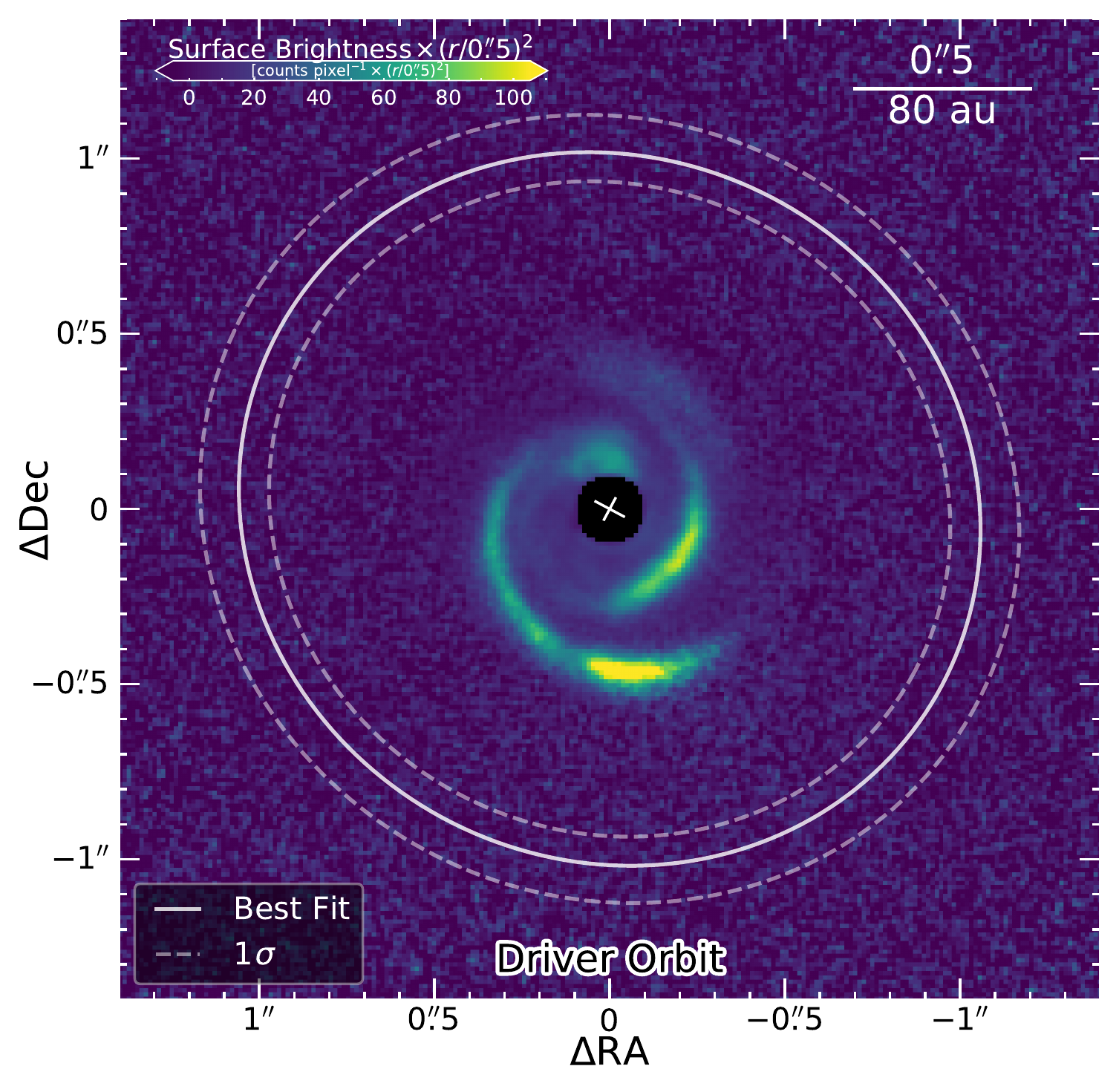}
\caption{Dynamically constrained, assumed circular orbit of a single arm driver. For a $1.9$ $M_\sun$ central star, the best-fit $\pm 1\sigma$ orbit has a radius of 
$1\farcs07_{-0\farcs09}^{+0\farcs11}$
($172_{-14}^{+18}$~au).
}
\label{fig_driver_location}
\end{figure}

In the companion driven scenario, the prominence and the symmetry of the two arms in the MWC 758 system suggest that they are produced by one companion of at least a few Jupiter masses \citep{fung15, dong17}. We thus fit the same pattern speed to both arms. The two spirals can be well fit by rigid body rotation at a rate of $0\fdg216\pm0\fdg016$~yr$^{-1}$. Taking into account of SPHERE's true north uncertainty, the updated motion rate is \begin{equation}
\omega = 0\fdg22\pm0\fdg03~{\rm yr}^{-1}.
\end{equation} 
This pattern speed points to a driver located at $172_{-14}^{+18}$~au, or $1\farcs07_{-0\farcs09}^{+0\farcs11}$, from the $1.9$ $M_\sun$ central star (Figure~\ref{fig_driver_location}). 

Our best fit measurement of the companion-driven spiral pattern speed is consistent with the \citetalias{ren18b} measurement within $3\sigma$,\footnote{Only $3\sigma$ uncertainties are well-constrained in \citetalias{ren18b}.}
while our derived uncertainty is ${\sim}40$ times smaller,
thanks to the use of the same instrument and the \Qphi\ maps that are the least biased by postprocessing methods.
In numerical simulations, a ${\sim}5$ Jupiter mass arm driver located at ${\sim}0\farcs9$ has been proposed by \citet{baruteau19}, which is within $2\sigma$ from our best fit companion location assuming a $1.9$ $M_\sun$ central star, or $1\sigma$ assuming a $1.56$ $M_\sun$ central star. 

\section{Discussion}

\subsection{Direct Imaging Constraints}
We obtain the direct imaging constraints on the mass of the putative planet orbiting MWC 758 with hot-start evolutionary models (i.e., Sonora, Bobcat; M.~Marley et al., in preparation) using $6.73$~h of Keck/NIRC2 $L'$-band archival observations: $2025$~s on 2015 October 24 \citep[Program ID: C220N2, PI: E.~Serabyn, ][]{reggiani18}, $11160$~s on 2016 February 12 (Program ID: U131N2, PI: E.~Chiang), $3200$~s on 2016 October 24 \citep[Program ID: C221N2, PI: G.~Ruane,][]{reggiani18}, and $7830$~s on 2017 February 02 (Program ID: U072, PI: E.~Chiang).  Following the method described in \citet{ruffio18} and taking into account of the orbital period uncertainty for the driver, we obtain an apparent $L$-band magnitude of $18$ at $99.9\%$ confidence level (i.e., a $3\sigma$ equivalent upper limit). Adopting an age of $10.9$~Myr \citep{garufi18}, this corresponds to a mass of $5~M_{\rm Jupiter}$.
%$5.2~M_{\rm Jupiter}$

Using the contrast curve of Keck/NIRC2 in $M_s$-band \citep{mawet19} and assuming Gaussian noise, we expect that a $5~M_{\rm Jupiter}$ planetary driver, whose mass has been predicted by \citet{dong15} and \citet{baruteau19} while with the semi-major axis updated in this study, can be detected at $5\sigma$ level if it is a hot-start planet using $4$ half-nights of NIRC2 $M_s$-band high contrast imaging observations (or at $3.5\sigma$ with the \citealp{spiegel12} cold-start model using the same observation).

\subsection{Eccentric Driver}
If the driver has a non-zero eccentricity $e$, its semi-major axis cannot be uniquely determined from an assessment of the instantaneous angular frequency due to the unknown orbital phase. Giant planets of several Jupiter masses interacting with a gaseous disk are expected to have their eccentricities quickly damped to below the disk aspect ratio, 
here about 20\% \citep{dunhill13, duffell15}.
Simulations have also shown that density waves excited by planets with $e\gtrsim0.2$ develop wiggles and bifurcations, as the waves launched at different phases interact \citep{li19, muley19}, which provide poor fits to the arms around MWC~758. A modest eccentricity introduces an uncertainty in the inferred planet location comparable to the uncertainty from pattern speed measurements -- for example, $e$ = 0.2 translates to a range of possible companion locations from $155$ to $190$~au. 

\subsection{Diverse Motion}
Noticing that the two arms could be excited by different companions or by different mechanisms \citep[e.g.,][]{forgan18}, we investigate their motion separately.

In the GI-induced scenario, the northern arm alone rotates at an angular speed of $0\fdg105\pm0\fdg013{\times}{(0.\!\!''5/r)^{3/2}}$~yr$^{-1}$, and the southern arm at $0\fdg025\pm0\fdg011{\times}{(0.\!\!''5/r)^{3/2}}$~yr$^{-1}$. A Keplerian disk around a $1.56$ $M_\sun$ central star  \citep{vioque18}
would be rotating at an angular speed of $0\fdg63{\times}(0.\!\!''5/r)^{3/2}$~yr$^{-1}$, which is a factor of ${>}5$ faster than the measurement and inconsistent with observations for both arms
at ${>}5\sigma$ levels.  Therefore, an even faster rotating Keplerian disk around a $1.9~M_\sun$ central star \citep{garufi18} is inconsistent with the motion rates. 

In the companion-driven scenario, if the arms are driven by different companions, we measure that the northern arm rotates by $0\fdg211\pm0\fdg019$ yr$^{-1}$, and the southern arm rotates by $0\fdg228\pm0\fdg027$ yr$^{-1}$. The two rates are within $1\sigma$ from each other, consistent with the expectation that the two arm are corotating and driven by the same driver.

\subsection{Model Selection}

From a statistical approach, our fitting results have a $\chi^2$ value of $3772$ in the single companion-induced scenario, and $3812$ in the global GI-induced scenario. Given that the two mechanisms are applied to the same number of data points (i.e., location pairs) and have the same number of variables (i.e., $1$ rotation speed variable and $2p+2=8$ polynomial coefficient variables), the Schwarz information criterion (SIC, \citealp{schwarz78}) difference is then $\Delta{\rm SIC}=\Delta\chi^2=40$, which is greater than $\Delta{\rm SIC}=10$ threshold for ``decisive'' evidence \citep{kass95} for model selection, making single planet driver the preferred mechanism. 

From another approach, assuming identical arm morphology between the two epochs, we can focus on the marginalized distribution for the speed parameter $\omega$ to quantify the difference. In this way, the above $\Delta\chi^2=40$ difference corresponds to a confidence level of $\sqrt{40}\sigma = 6.3\sigma$ (Chapter 15.6 of \citealp{press92}), which makes the single planet driver mechanism to be more consistent with our observations. Similarly, we apply the above analysis to each individual arm and find that the planet driven scenario is more consistent with the observations. 

\subsection{Robustness Estimation} 
In our analysis, we have investigated polynomials up to $p = 7$ degrees. We obtain the lowest SIC that penalizes excessive use of parameters at the cubic form when $p =3$. When $p\geq3$, we observe no discernible best fit angular speeds, and thus we use the cubic description of the spiral arms in our analysis. To robustly obtain the best-fit and uncertainty for these parameters, we have investigated the impact from different chi-squared minimization methods, including {\tt scipy.optimize.curve\_fit} \citep{scipy19} and orthogonal least squares fitting code {\tt scipy.odr} \citep{odr89}, and no discernible difference was obtained. We  present in this study the results from {\tt scipy.optimize.curve\_fit}.

The flaring of the disk \citep[e.g.,][]{stolker16b, rosotti20} does not bias our estimation. We use {\tt diskmap} \citep{stolker16b} to deproject the disk images with various flaring exponents (i.e., 0, 0.5, 1, 1.1, 1.2) and repeat the motion measurement, the results are all consistent within $1\sigma$. In addition, we randomly varied the inclination and position angle for the disk within ${\pm}5^\circ$ for $10^3$ times, and the motion rates for the single planet driver are $0\fdg22\pm0\fdg06$~yr$^{-1}$ and being consistent with the original estimate within $2\sigma$.

Our measurements are not biased by star centering uncertainties in two aspects. First, we use the same pipeline (i.e., {\tt IRDAP}) with identical reduction parameters to minimize systematic offset. Second, even if there are offsets, the inclination of the disk would impact in the individual arm rotation rates by returning different angular speed measurements in the planet-driven scenario, which are indistinguishable since our measurements are within $1\sigma$.

The impact from individual location pairs is negligible. We experiment by randomly discarding up to $25\%$ of the pairs and repeating the speed measurement procedure for $10^4$ times. The best-fit rotation rate for a single driver is found to be $0\fdg22\pm0\fdg02$~yr$^{-1}$, consistent with our initial measurement to within $1\sigma$.

The morphology of the spiral arms is consistent with being circular when the stellocentric separation is less than $40$~au ($0\farcs25$, Figure~\ref{fig-measure}), and such regions have a seemingly outward motion in Figure~\ref{fig-measure}. We argue here that this does not bias our results. When we ignore these location pairs and repeat our fitting, the results do not change by more than $1\sigma$. In fact, since the number of data points in these regions is less than $25\%$ of the total number of data points, this scenario has been investigated in the above procedure of random location pair rejection.

\subsection{Possible Systematics}

The 2015 March data are taken in field-tracking mode with an non-ideal HWP control law\footnote{See SPHERE User Manual at \url{http://www.eso.org/sci/facilities/paranal/instruments/sphere/doc.html}}, which has been rectified in 2015 late April. The non-ideal control law causes the polarization direction to rotate on the detector during the observations, which we correct for using the Mueller matrix model of {\tt IRDAP}. We confirm the proper correction of the images with {\tt IRDAP} by comparing the uncorrected and corrected polarimetric images cubes for both the 2015 and 2019 data.

The 2015 data are taken with a sub-optimal order of HWP switch angles ($0^\circ$, $22\fdg5$, $45^\circ$, $67\fdg5$ instead of $0^\circ$, $45^\circ$, $22\fdg5$, $67\fdg5$ for the 2019 data) and with a high number of integrations per HWP position (6 instead of 2 for the 2019 data). As a result, for the 2015 data a measurement of $Q$ or $U$ lasts approximately 10~min 10~s, compared to 2~min 25~s for the 2019 data. Because the polarization direction rotates on the detector during the 2015 measurements, there may be a global impact on the final \Qphi~image. The 2019 \Qphi\ image may also be slightly affected, because in the reduction of pupil-tracking data the images are derotated after computing the double difference \citep{vanholstein17}. Given that the disk has a non-zero inclination and position angle, these effects could bias the speed measurement when we deproject the image to face-on views. Therefore, it should be reflected in the measured individual rotation speeds under the planet driver scenario that calculates global offsets. Nevertheless, since the individual arm rotation rates are consistent within $1\sigma$, we do not expect the global impact from observation strategy 
to 
bias our results at more than $1\sigma$ level.

There are caveats in our measurements. Finite inclinations are known to produce distortions in images that are hard to correct in deprojection \citep{dong16b}. However, this effect usually becomes prominent only at inclinations larger than ${\sim}20^\circ$, and we do not expect strong morphology distortions in the MWC 758 disk whose inclination is ${\sim}20^\circ$. In addition, perturbations from a theorized inner companion in the disk \citep{baruteau19} may cause slight changes to the shape of the spiral arms that are unrelated to their primary driver, thus affecting our pattern speed measurements. Furthermore, the change in illumination may slightly change the observed features on disk surface \citep{montesinos16}. Future multi-epoch observations of MWC~758 are necessary to quantify such effects.

\section{Summary}

We have established a 5~yr baseline and obtained the most accurate pattern speed measurement of spiral arms in a protoplanetary disk to date. For the two prominent spiral arms surrounding MWC~758, we witness global scale shadowing effects and measure the motion between the two epochs to test their formation and motion mechanisms.

We found that the measured motion of spirals disfavors their GI origin. This is the first time that it has been shown for any protoplanetary disk. Meanwhile, our motion analysis suggests a single planet driving both spiral arms. For a $1.9~M_\sun$ central star, our measurement pinpoints a semi-major axis of $172_{-14}^{+18}$~au for the planet driver if its orbit is circular. Using archival Keck/NIRC2 $L'$-band observations totaling $6.73$~hr, we obtain a $3\sigma$-equivalent upper limit of $5~M_{\rm Jupiter}$ for the location of this driver using hot start planet formation models. The inferred spiral arm driver in the MWC 758 system is ideal for Keck/NIRC2, VLT/Enhanced Resolution Imager and Spectrograph (ERIS), and \textit{James Webb Space Telescope} direct detections in longer wavelengths, and for Atacama Large Millimeter/submillimeter Array (ALMA) circumplanetary disk exploration.

\facilities{Very Large Telescope (SPHERE), Keck:II (NIRC2)}

\software{{\tt diskmap} \citep{stolker16b}, {\tt IRDAP} \citep{vanholstein17, vanholstein20}, {\tt scipy} \citep{scipy19}}

\acknowledgements 
We thank the anonymous referee for comments that improved the clarity of this Letter, and Cassandra Hall for useful discussions. Based on observations collected at the European Organisation for Astronomical Research in the Southern Hemisphere under ESO programmes 060.A-9389(A) and 104.C-0472(A). B.R.~thanks Christian Ginski for discussions on shadowing effects, R\'emi Soummer for initiating the Archival Legacy Investigations of Circumstellar Environments (ALICE) project that set up the stage for \citetalias{ren18b} and this Letter. T.E.~was supported in part by NASA Grants NNX15AD95G/NEXSS, NNX15AC89G, and NSF AST-1518332. A.L.M.~acknowledges the financial support of the F.R.S.-FNRS through a postdoctoral researcher grant. Some of the data presented herein were obtained at the W.~M.~Keck Observatory, which is operated as a scientific partnership among the California Institute of Technology, the University of California and the National Aeronautics and Space Administration. The Observatory was made possible by the generous financial support of the W.~M.~Keck Foundation. The authors wish to recognize and acknowledge the very significant cultural role and reverence that the summit of Maunakea has always had within the indigenous Hawaiian community.  We are most fortunate to have the opportunity to conduct observations from this mountain.

\appendix
\section{Spiral Arm Motion}\label{math-measure}
To constrain spiral arm morphology and motion, we note that $\theta$ was expressed as function of $r$ to allow for matrix inversion using linear algebra in \citetalias{ren18b}. After inspecting the stability of the high resolution \Qphi\ images in this study, here we switch their relationship in order to allow for precise measurement of $r$ as a function of $\theta$. 

For a rotating spiral arm, a $(\theta_i, r_i)$ pair will be updated to $(\theta_i + \omega_{{\rm model}, i}t, r_i)$ at a new epoch, where $t$ is the temporal separation between the two observations, and $\omega_{{\rm model}, i}$ the angular speed in the scenario that is either companion-driven (``comp'') or gravity instability--induced (``GI''). For a total of $E$ epochs, we describe the location pairs using polynomials with $(E-1)$ dummy variables, \begin{equation}
r_i^{\rm (model)}(\theta_i) = \sum_{j = 0}^p c_j \left(\theta_i + \omega_{{\rm model}, i} \sum_{k=2}^{E} t_k D_k\right)^j,
\label{eq-pair-model}
\end{equation}
where $p\in\mathbb{N}$ describes the degree of the polynomial at the first epoch when $r_i(\theta_i) = \sum_{j=0}^pc_j\theta_i^j$ and $c_j \in \mathbb{R}$ is the coefficient for the $j$-th power term, $t_k \in \mathbb{R}$ is temporal separation between epoch $k$ and $1$, and $D_k \in \{0, 1\}$ is a dummy variable that equals $1$ only when the $(\theta_i, r_i)$ pair is obtained at epoch $k$. We note that the above equation is to describe the location pairs using $E$ polynomials that are mutually related through angular offsets.

For a total of $m$ measured location pairs, we minimize the following chi-squared statistic,
\begin{equation}
\chi^{2} = \sum_{i=1}^m \left(\frac{r_i - r_i^{\rm (model)}(\theta_i)}{\delta r_i}\right)^2,\label{eq-chi2}
\end{equation}
to obtain the motion rate. In this way, we can simultaneously constrain the morphological parameters and motion rate using all available location pairs.

\subsection{Companion Driven}
If a spiral arm is driven by a companion a companion on a circular orbit \citep[``comp'': ][]{kley12, dong15, zhu15, bae16} that is located at a stellocentric position of $r_{\rm comp}$, the entire arm corotates with the companion at the Keplerian angular speed of the companion, $\omega_{\rm comp}$. In this way, an arm observed at different epochs is shifted in the azimuthal direction while maintaining its shape in the disk plane. A $(\theta_i, r_i)$ pair will be updated to $(\theta_i + \omega_{\rm comp}t, r_i)$ at a new epoch. Equation~\eqref{eq-pair-model} then becomes
\begin{equation}
r_i^{\rm (comp)}(\theta_i) = \sum_{j = 0}^p c_j \left(\theta_i + \omega_{\rm comp} \sum_{k=2}^{E} t_k D_k\right)^j.\label{eq-pair-pd}
\end{equation}
We note that the physical meaning of the above equation is to fit offseted identical polynomials to the data.

In the companion driven scenario, we substitute Equation~\eqref{eq-pair-pd} into Equation~\eqref{eq-chi2}, 
i.e.,
\begin{equation}
\chi^{2{\rm (comp)}} 	= \sum_{i=1}^m \left[\frac{r_i - \sum_{j = 0}^p c_j \left(\theta_i + \omega_{\rm comp} \sum_{k=2}^{E} t_k D_k\right)^j}{\delta r_i} \right]^2, \label{eq-chi2-pd}
\end{equation}
to obtain the motion rate when the arms co-move.

When there are a total of $s$ spiral arms, we denote their arm location measurements with $(r_{i, l}, \theta_{i, l})$ for $l \in \{1, \cdots, s\}$. These spiral arms rotate at the same rate if all of them are driven by the same driver, then Equation~\eqref{eq-pair-pd} becomes
\begin{equation}
r_{i, l}^{\rm (comp)}(\theta_{i, l}) = \sum_{l = 1}^s \sum_{j = 0}^p c_{j, l} \left(\theta_{i, l} + \omega_{\rm comp} \sum_{k=2}^{E} t_k D_k\right)^j D_l,\label{eq-pair-pd-multi}
\end{equation}
where $D_l \in \{0, 1\}$ is a dummy variable that equals $1$ only when the $(\theta_{i, l}, r_{i, l})$ pair is obtained from spiral arm $l$. The corresponding $\chi^2$ minimization formula is obtained by substituting the $r$ expressions in Equation~\eqref{eq-pair-pd-multi} to the $\chi^2$ expression in Equation~\eqref{eq-chi2}.  The physical meaning of the above equation is to fit multiple arms using Equation~\eqref{eq-pair-pd} but with a constraint that their motion rates are identical.

\subsection{GI-induced}
If a spiral arm is excited by GI \citep{lodatorice15, dong15b, kratter16}, each part of the arm rotates at its local Keplerian angular speeds in the disk plane on timescale much smaller than the local dynamical timescale (spiral arms disappear and reemerge on longer timescale).
For any $(\theta_i, r_i)$ location pair, its location at a new epoch will be $\left(\theta_i + \frac{r_0^{3/2}}{r_i^{3/2}}\omega_{0}t, r_i\right)$, where $\omega_{0}$
is the Keplerian angular speed at stellocentric separation $r_0$. In this scenario, Equations~\eqref{eq-pair-model} and \eqref{eq-chi2} have a power law attenuation in their angular speed terms, i.e.,

\begin{equation}
r_i^{\rm (GI)}(\theta_i) = \sum_{j = 0}^p c_j \left(\theta_i + \omega_{0}\frac{r_0^{3/2}}{r_i^{3/2}} \sum_{k=2}^{E} t_k D_k\right)^j,\label{eq-pair-gm}
\end{equation}
and
\begin{equation}
\chi^{2{\rm (GI)}} = \sum_{i=1}^m \left[\frac{r_i - \sum_{j = 0}^p c_j \left(\theta_i + \omega_{0}\frac{r_0^{3/2}}{r_i^{3/2}} \sum_{k=2}^{E} t_k D_k\right)^j}{\delta r_i} \right]^2,\label{eq-chi2-gm}
\end{equation}
respectively. We note that the physical meaning of the above two equations is to fit lines that have $r^{-3/2}$-dependent angular offsets.

Similarly, for a total of $s$ spiral arms undergoing the same local Keplerian motion, the corresponding power law attenuation form of Equation~\eqref{eq-pair-pd-multi} is
%\begin{linenomath*}
\begin{equation}
r_{i, l}^{\rm (GI)}(\theta_{i, l})  = \sum_{l = 1}^s\sum_{j = 0}^p c_{j, l} \left(\theta_{i, l} + \omega_{0}\frac{r_0^{3/2}}{r_i^{3/2}} \sum_{k=2}^{E} t_k D_k\right)^j D_l.\label{eq-pair-gm-multi}
\end{equation}
%\end{linenomath*}
 The corresponding $\chi^2$ minimization formula is obtained by substituting the $r$ expression in Equation~\eqref{eq-pair-gm-multi} to the $\chi^2$ expression in Equation~\eqref{eq-chi2}. The physical meaning of the above equation is similar to Equation~\eqref{eq-pair-pd-multi} but for the GI-induced mechanism.
 
In this study, we have $E=2$ epochs with a temporal separation of $t=4.71$~yr. We report the motion with polynomial degree $p=3$ for both scenarios. 

\bibliography{refs}
\end{CJK*}
\end{document}